\documentclass[prl,twocolumn,floatfix]{revtex4}

\usepackage{amsfonts}
\usepackage{amsmath}
\usepackage{verbatim}
\usepackage[dvips]{graphicx}
\usepackage{subfigure}

\def\Tr{\textrm}
\def\dd{\textrm{d}}

\def\vv{\Tr{v}}
\def\pp{\Tr{p}}

\begin{document}

\title{Charge and spin fractionalization in strongly correlated topological insulators}
\author{Predrag Nikoli\'c$^{1,2}$}
\affiliation{$^1$School of Physics, Astronomy and Computational Sciences, George Mason University, Fairfax, VA 22030, USA}
\affiliation{$^2$Institute for Quantum Matter at Johns Hopkins University, Baltimore, MD 21218, USA}
\date{\today}

\begin{abstract}

We construct an effective topological Landau-Ginzburg theory that describes general SU(2) incompressible quantum liquids of strongly correlated particles in two spatial dimensions. This theory characterizes the fractionalization of quasiparticle quantum numbers and statistics in relation to the topological ground-state symmetries, and generalizes the Chern-Simons, BF and hierarchical effective gauge theories to an arbitrary representation of the SU(2) symmetry group. Our main focus are fractional topological insulators with time-reversal symmetry, which are treated as SU(2) generalizations of the quantum Hall effect.

\end{abstract}

\maketitle

\section{Introduction}

A two dimensional electron gas in a strong magnetic field can exhibit fractional quantum Hall effect (FQHE) and fluctuations that carry a fractionally quantized amount of the electron's elementary charge \cite{WenQFT2004}. Similar fractionalization is also possible in topological insulators (TIs) with time-reversal (TR) symmetry \cite{Hasan2010, Qi2010a, Moore2010}. Owing to a strong spin-orbit coupling, these materials exhibit protected gapless modes at their boundaries, but differ from the quantum Hall systems by their spectra and by lacking a conserved ``charge'' (spin). The latter prevents the observation of a quantized transverse conductivity and obscures the fact that the topological features of TIs are born in the SU(2) extension of the quantum Hall effect \cite{Frohlich1992}.

This paper is motivated by the growing interest in strongly correlated TIs that feature fractional excitations \cite{Qi2008b, Levin2009, Karch2010, Cho2010, Maciejko2010, Swingle2011, Neupert2011}, and their promise to enable novel topological orders that have no analogue in quantum Hall states. Our goal is to construct an effective field theory of generic 2D fractional TIs that can capture the spin non-conserving character of spin-orbit couplings.

Fractional TIs can be obtained in various systems, and likely will be observed in the foreseeable future. One promising system are ultra-cold gases of bosonic atoms trapped in quasi 2D optical lattices. Superfluid to Mott insulator transitions can be easily arranged to remove any energy scales that could compete with the spin-orbit coupling \cite{Greiner2002}. The ensuing quantum critical point (QCP) is therefore highly sensitive to relevant perturbations such as the spin-orbit coupling. The recent development of artificial gauge fields for neutral atoms, created by stimulated Raman transitions between internal atomic states, has introduced effective spin-orbit couplings in atomic gases\cite{Lin2011, Sau2011}, and even promises the ability to reach the fractional quantum Hall regime \cite{Campbell2011, Cooper2011}. Bosonic atoms in such spin-orbit fields are likely to have stable fractional TI phases in the critical fan of their Mott QCP.

Promising solid-state systems include spin-orbit materials with significant Coulomb interactions, and TI-superconductor heterostructures. The former materials feature frustration of the electrons' kinetic energy, either by lattice geometry  \cite{Pesin2010} or gauge fields \cite{Sun2011}. Low-energy excitations can carry spin in these systems, so their coupling to orbital motion can result with non-trivial topological properties.

The heterostructure systems rely on a superconducting material to induce correlations among electrons in the proximate quantum well made from a band-insulating TI such as Bi$_2$Se$_3$ or Bi$_2$Te$_3$ \cite{Nikolic2010b, Nikolic2012a}. The proximity effect produces Cooper pairs in the TI, which in turn can form a superconducting state, or a bosonic Mott-insulator \cite{Nikolic2010b} enabled by the surface hybridization bandgap in the TI quantum well. A phase transition between these states in the quantum well can be tuned by a gate voltage, yielding a similar QCP to that mentioned in the context of bosonic atoms above. The crucial ingredient here are inter-orbital $p$-wave Cooper pairs, which are allowed in the TI quantum well and dynamically enhanced by their spin-orbit coupling. The superconducting phase of such $p$-wave Cooper pairs is expected to host a TR-invariant vortex lattice due to the spin-orbit SU(2) flux. The quantum melting of this vortex lattice induced by a gate voltage is expected to produce an incompressible quantum liquid that could naturally be a fractional TI. The proximity effect in TI \emph{quantum wells} outlined here should be contrasted with that in \emph{bulk three-dimensional} TIs, which provides a route to Majorana quasiparticles \cite{Fu2008} but cannot give rise to incompressible topological quantum liquids.

The simplest and most universal theoretical tool for describing fractional topological states of matter is the effective Chern-Simons (CS) gauge theory \cite{WenQFT2004}. Its action is stationary when the (spin) Hall conductivity is properly quantized. This feature provides the essential justification of the CS theory whose microscopic derivation is not available. However, it also makes the application of CS theories to fractional TIs problematic. The Rashba spin-orbit coupling produces TIs without a spin-Hall effect because it does not conserve spin. Thus, we will seek here a generalization of the CS effective theory to incompressible quantum liquids without spin conservation.

We will first introduce an SU(2) gauge-theory description of the spin-orbit coupling in TIs \cite{Frohlich1992}. It has the advantage of relying on the same physical picture used in the studies of FQHE, while being able to naturally handle the spin non-conservation. The SU(2) gauge symmetry is approximate and various perturbations are allowed to remove it in realistic systems. Fortunately, this cannot jeopardize the topologically protected many-body quantum entanglement in the two-dimensional bulk, and its manifestations such as fractional statistics that the SU(2) formalism will capture. We will first develop the SU(2) theory in band-insulating TIs, and then use it to construct the field theory of fractional ground states that could be stabilized in the presence of appropriate interactions as surveyed before. While mainly focused on the theory construction, we will make physical predictions along the way on the possible fractional quantum numbers of excitations in relation to the symmetries and external magnetic and spin-orbit fields.

The topological effective field theory that we will construct has a general form that can be applied to quantum Hall effects of arbitrary SU($N$) symmetry groups. The U(1) Hall effect is created by magnetic fields, while the SU(2) Hall effect can be obtained by spin-orbit couplings that conserve a spin projection. The symmetry of the model will also enable describing more general SU(2) topological states without spin conservation, created for example by the Rashba spin-orbit coupling. Most importantly, being a generalization of the CS theory, our model can naturally address strongly correlated incompressible quantum liquids with fractional quasiparticles. This is the main motivation for its construction.

\section{Single-particle topological dynamics}

Consider the following minimal model of a band-insulating TI quantum well, where electrons have the two-state spin $\sigma^z$ and orbital $\tau^z$ quantum numbers:
\begin{equation}\label{Bernevig}
H = \frac{({\bf p}-\tau^z \boldsymbol{\mathcal{A}})^{2}}{2m} + \Delta\tau^{x} - \mu
  \quad , \quad \boldsymbol{\mathcal{A}}=-mv(\hat{{\bf z}}\times{\bf S}) \ .
\end{equation}
${\bf S} = \frac{1}{2} \sigma^a \hat{\bf r}^a$ is the vector spin operator, $\sigma^a$ and $\tau^a$ are Pauli matrices that operate on the spin and orbital degrees of freedom respectively, and we set $\hbar=1$. This Hamiltonian describes an electron moving in the external SU(2) gauge field $\boldsymbol{\mathcal{A}}$ whose SU(2) charge $g=\tau^z$ depends on its orbital state. The given form of $\boldsymbol{\mathcal{A}}$ reproduces the Rashba spin-orbit coupling $H_{\textrm{so}} = v\,\hat{{\bf z}} ({\bf S}\times{\bf p}) \tau^{z}$, and the orbital index $\tau^z$ can be interpreted as the top or bottom surface of the quantum well. Tunneling $\Delta$ between the two surfaces opens a band-gap in an otherwise massless Dirac spectrum, provided that we interpret this model as being applicable only below a cut-off momentum scale $\Lambda = \sqrt{(mv)^2-(\Delta/v)^2}$ (such a momentum cut-off is given by the lattice constant in realistic systems). The spin-orbit coupling $v$, however, is responsible for all topological properties, which is clearly revealed by the non-zero ``magnetic'' SU(2) flux density:
\begin{equation}
\Phi^\mu = \epsilon^{\mu\nu\lambda} ( \partial_\nu \mathcal{A}_\lambda - ig \mathcal{A}_\nu \mathcal{A}_\lambda )
         = \frac{1}{2}(mv)^{2}\delta_{\mu 0}\,\tau^z\sigma^z
\end{equation}
We have used Einstein's notation for summation over the repeated indices and the (2+1)D Levi-Civita tensor $\epsilon^{\mu\nu\lambda}$ ($\mu,\nu\dots$ are space-time directions, $i,j,\dots$ are spatial directions, $\mu=0$ indicates the temporal direction). This model naturally describes the hybridized ``surface'' spectrum of the Bi$_2$Te$_3$, Bi$_2$Se$_3$ and other TI quantum wells \cite{Zhang2009b}. It is also related to the model of HgTe quantum wells with band inversion \cite{Bernevig2006} by a change of representation, although its spin transport properties are fundamentally different (a perfect HgTe crystal with non-interacting electrons would exhibit quantum spin-Hall effect at low energies).

\begin{table*}
\centering
\begin{tabular}{|c||c|c|c|}
\hline 
 & U(1)$\times$SU(2) & charge U(1) & spin SU(2) \\
\hline\hline 
group generators & $1,\gamma^{a} \;\; \small\textrm{(m)}$ & $1 \;\; \small\textrm{(m)}$ & $\gamma^{a} \;\;\small\textrm{(m)}$ \\
\hline 
external gauge field & $\mathcal{A}_{\mu}^{\phantom{a}}=a_{\mu}^{\phantom{a}}1+A_{\mu}^{a}\gamma^{a} \;\; \small\textrm{(m)}$
 & $a_{\mu}^{\phantom{a}} \;\; \small\textrm{(s)}$ & $A_{\mu}^{a} \;\; \small\textrm{(s)}$ \\
\hline 
external gauge flux & $\Phi_{\mu}^{\phantom{a}}=\phi_{\mu}^{\phantom{a}}1+\Phi_{\mu}^{a}\gamma^{a} \;\; \small\textrm{(m)}$
 & $\phi_{\mu}^{\phantom{a}} \;\; \small\textrm{(s)}$ & $\Phi_{\mu}^{a} \;\; \small\textrm{(s)}$
\\
\hline 
particle current operator &  & $j_{\mu}^{\phantom{a}} \;\; \small\textrm{(m)}$ & $J_{\mu}^{a} \;\; \small\textrm{(m)}$ \\
\hline 
particle current density &  & $j_{\textrm{p}\mu}^{\phantom{a}} \;\; \small\textrm{(s)}$ & $J_{\textrm{p}\mu}^{a} \;\; \small\textrm{(s)}$ \\
\hline 
vortex current density &  & $j_{\textrm{v}\mu}^{\phantom{a}} \;\; \small\textrm{(s)}$ & $J_{\textrm{v}\mu}^{a} \;\; \small\textrm{(s)}$ \\
\hline
\end{tabular}
\caption{\label{SymbolRef}The reference for frequently used symbols. All shown quantities are (2+1)D space-time vectors indexed by $\mu \in \lbrace 0,x,y \rbrace$, where $0$ indicates the time direction. The superscript $a \in \lbrace x,y,z \rbrace$ labels the three SU(2) generators $\gamma^a$, which are $(2S+1)$-dimensional matrices in any spin-$S$ representation (Pauli matrices if the particles have spin $S=\frac{1}{2}$). Any repeated index is summed over, and (s) or (m) indicate if the symbol is used as a scalar or a matrix respectively.}
\end{table*}

Knowing the gauge-field description of the spin-orbit coupling, we can generalize the model (\ref{Bernevig}) to any combination of spin-orbit couplings and external electromagnetic fields acting on particles with arbitrary spin $S$. This will enable us to eventually apply the theory not only to uncorrelated topological band-insulators, but also to correlated states of composite particles such as Cooper pairs in solid-state materials, or bosonic atoms in ultracold atom experiments. The model (\ref{Bernevig}) will serve from now on only as a guide for the type of the gauge field and couplings that we must choose to describe electromagnetic fields or spin-orbit interactions. The general U(1)$\times$SU(2) gauge field is the matrix $\mathcal{A}_\mu = a_\mu + A_\mu^a \gamma^a$, where $\gamma^a$ for $a\in\lbrace x,y,z \rbrace$ are the SU(2) generators in the spin $S$ representation. The Table \ref{SymbolRef} summarizes the use of various symbols in the rest of the paper. The gauge flux
\begin{equation}
\Phi^\mu = \epsilon^{\mu\nu\lambda} ( \partial_\nu \mathcal{A}_\lambda - ig \mathcal{A}_\nu \mathcal{A}_\lambda )
  = \phi_\mu^{\phantom{a}} + \Phi_\mu^a \gamma^a
\end{equation}
couples to the U(1) charge $e$ and SU(2) charge $g$ in the Hamiltonian
\begin{equation}\label{Hamiltonian}
H_0 = \frac{({\bf p} - e {\bf a} - g {\bf A}^a \gamma^a)^2}{2m} - ea_0 - gA_0^a\gamma^a + \cdots \ ,
\end{equation}
where the dots denote any other terms that operate in the orbital space and provide an energy gap. The temporal $\Phi^0$ and spatial $\Phi^i$ fluxes correspond to ``magnetic'' and $90^o$-rotated ``electric'' fields respectively. Defining the charge $j_\mu$ and spin $J_\mu^a$ current operators,
\begin{eqnarray}\label{CurrentOperators}
  j_{0}=1                 & \quad,\quad &    j_{i}=\frac{1}{m}(p_{i}-ea_{i}-gA_{i}^a\gamma^a) \nonumber \\
  J_{0}^{a}=\gamma^{a}    & \quad,\quad &    J_{i}^{a}=\frac{1}{2}\lbrace\gamma^{a},j_{i}\rbrace
\end{eqnarray}
where braces denote an anti-commutator, we obtain the following Heisenberg equation of motion from (\ref{Hamiltonian}):
\begin{equation}\label{Heisenberg}
\frac{\dd j_{i}}{\dd t} = i \lbrack H_0,j_i \rbrack = \frac{e}{m}\epsilon_{i\nu\lambda}j_{\nu}\phi_{\lambda}
  + \frac{g}{2m}\epsilon_{i\nu\lambda}\lbrace j_{\nu},\Phi_{\lambda}^a\gamma^a\rbrace \ .
\end{equation}
If the gauge field $\mathcal{A}_\mu$ and flux $\Phi_\mu$ matrices commute with the Hamiltonian, the general periodic solution is
\begin{equation}
j_{i}(t) = c_{i}e^{i\omega t}e^{i\gamma^{a}\omega^{a}t} + \delta j_i \ ,
\end{equation}
where the first term describes cyclotron motion with $\omega = e\phi_0/m$, $\omega^a = g\Phi_0^a/m$ and $c_y = i c_x$. The second term $\delta j_i$ is a constant drift current perpendicular to both ``electric'' and ``magnetic'' fields.

We will now concentrate on the drift current kinematics in quantum spin-Hall states whose global spin U(1) symmetry is reflected by $\lbrack \Phi_\mu, H_0 \rbrack=0$. Then, we are free to work in the representation $\Phi_\mu = \phi_\mu + \Phi_\mu^z \gamma^z$, where $\gamma^z$ is the standard diagonal angular momentum matrix in the spin $S$ representation. Setting $\dd j_i / \dd t = 0$ in (\ref{Heisenberg}), we find:
\begin{equation}\label{dJ}
\delta j_{i}^{\phantom{z}}=\Bigl(e\phi_{0}^{\phantom{z}}+g\Phi_{0}^z\gamma^{z}\Bigr)^{-1}
       \Bigl(e\phi_{i}^{\phantom{z}}+g\Phi_{i}^z\gamma^{z}\Bigr) \ .
\end{equation}
What we will need, however, is a slightly different formula
\begin{eqnarray}\label{EqMot}
\delta j_{i}^{\phantom{z}} &\!\!\!=\!\!\!&
    \Bigl\lbrack e^{2}\phi_{0}^{2}+2eg\phi_{0}\Phi_{0}\gamma^{z}+g^{2}\Phi_{0}^{2}(\gamma^{z})^{2}\Bigl\rbrack^{-1} \\
 && \Bigl\lbrack e^{2}\phi_{0}\phi_{i}
    +eg(\phi_{0}\Phi_{i}+\phi_{i}\Phi_{0})\gamma^{z}
    +g^{2}\Phi_{0}\Phi_{i}(\gamma^{z})^{2}\Bigr\rbrack \ , \nonumber
\end{eqnarray}
which is obtained by inserting $e\phi_{0}^{\phantom{z}} + g\Phi_{0}^z\gamma^{z}$ and its inverse in the solution for $\delta j_i$ in (\ref{dJ}).

The equation (\ref{dJ}) and its equivalent (\ref{EqMot}) describe quantum Hall effects. Consider applying an external electric field $E_i = -\partial_i a_0 -\partial_0 a_i$ in the quantum well plane. The corresponding electric flux is $\phi_i = \epsilon^{i\mu\nu} \partial_\mu a_\nu = \epsilon_{ij} E_j$ (where $\epsilon_{ij} \equiv \epsilon_{0ij}$), so its orientation in space is perpendicular to that of the electric field vector. We will assume that there are no spin-dependent electric fields, $\Phi_i^a = 0$. Then, a U(1) magnetic field $B$ perpendicular to the plane, whose corresponding magnetic flux is $B \equiv \phi_0 = \epsilon^{0ij} \partial_i a_j$, gives rise to a steady current per particle $\delta j_i = \epsilon_{ij} E_j / B$ according to (\ref{dJ}). If the particle and flux-quantum densities are $n$ and $n_\phi$ respectively, then the charge current density $n e \, \delta j_i = \sigma_{xy} \epsilon_{ij} E_j$ written in the standard units is determined by the Hall conductivity $\sigma_{xy}  = \nu e^2 / h$, where $\nu = n/n_\phi$ is the filling factor and $B = n_\phi h/e$. Therefore, (\ref{dJ}) captures the quantum Hall effect. Of course, no interactions are included in (\ref{dJ}), so it can adequately describe only integer quantum Hall effects. A similar relationship of (\ref{dJ}) to the non-interacting quantum spin-Hall effect can be obtained when electrons experience an $S^z$-conserving spin-orbit coupling with the SU(2) flux $\Phi_0^z$ instead of a magnetic field. The SU(2) gauge field would have to be $\mathcal{A}_\mu = A_\mu \gamma^z$ in some gauge, where $A_\mu$ is a U(1) gauge field satisfying $\epsilon^{\mu\nu\lambda} \partial_\nu^{\phantom{z}} A_\lambda^{\phantom{z}} = \Phi_0^z \delta_{\mu 0}^{\phantom{z}}$. Recall, however, that the spin-orbit gauge field in (\ref{Bernevig}) does not conserve $S^z$ and leads to a TI without a spin-Hall effect.

\section{Many-body topological dynamics}

We now turn to interacting systems and construct an effective topological field theory of spin $S$ particles in quantum Hall states shaped by magnetic fields, spin-orbit couplings, or both. Such a theory cannot be derived microscopically, but will be justified in the description of universal phenomena on the basis of symmetries, as is well established in the theory of critical phenomena. Our approach to constructing the field theory is equivalent to that of high energy physics where the microscopic theory (at cut-off scales) is not known and a Lagrangian is simply constructed. The field theory Lagrangian must contain the minimal set of degrees of freedom that can capture the dynamics of the given system, and the simplest couplings among them allowed by symmetries (relevant in the renormalization group sense). It must also reproduce the known classical equations of motion at its saddle point via the stationary action principle. The field-theoretical second quantization automatically satisfies these requirements when the classical dynamics is known. For our purposes, we will treat (\ref{EqMot}) as the classical equation of motion that the field theory must reproduce. More accurately, we will rely on symmetries to formulate a standard Landau-Ginzburg theory of generic spin $S$ particles in two spatial dimensions, and add to it a topological term that satisfies all symmetry requirements and whose saddle-point is equivalent to (\ref{EqMot}). The Landau-Ginzburg part is responsible for the dynamics, while the topological term merely specifies the classically observable kinematics of drift currents in quantum Hall states. In fact, the topological term will be relevant only in the quantum Hall phases where the Landau-Ginzburg dynamics keeps both particles and vortices mobile. After the construction is done, we will identify the fields as being vortex rather than physical particle degrees of freedom. We emphasize that the CS theory of charge quantum Hall effects is constructed using the same principle as a Lagrangian of abstract gauge fields whose saddle-point describes the quantized Hall conductivity and incompressible density \cite{WenQFT2004}.

We will set $e=g=1$ for simplicity. Consider a Lagrangian density $\mathcal{L} = \mathcal{L}_\textrm{LG} + \mathcal{L}_t$ in imaginary time:
\begin{eqnarray}\label{TopLG}
&& \!\!\!\! \mathcal{L}_\textrm{LG} = \frac{K}{2}\Bigl\vert(\partial_{\mu}-i\mathcal{A}_{\mu})\psi\Bigr\vert^2
       -t|\psi|^2-t'\psi^\dagger\Phi_0\psi \\
&& ~~~ +u|\psi|^4+v|\psi^{\dagger}\gamma^{a}\psi|^2
       +v'|\psi^{\dagger}\Phi_{0}\psi|^2 \nonumber \\[0.1in]
&& \!\!\!\! \mathcal{L}_t = -\frac{i\eta}{2} \psi^{\dagger}\epsilon^{\mu\nu\lambda}\Bigl\lbrack
     (\partial_{\mu}-i\mathcal{A}_{\mu})
     \Bigl\lbrace\partial_{\nu}-i\mathcal{A}_{\nu},\Phi_{0}\Bigr\rbrace
     (\partial_{\lambda}-i\mathcal{A}_{\lambda}) \nonumber \\
&& ~~~ +\Bigl\lbrace(\partial_{\mu}-i\mathcal{A}_{\mu})(\partial_{\nu}-i\mathcal{A}_{\nu})
     (\partial_{\lambda}-i\mathcal{A}_{\lambda}),\Phi_{0}\Bigr\rbrace\Bigr\rbrack\psi \ , \nonumber
\end{eqnarray}
where $\mathcal{L}_\textrm{LG}$ is the Landau-Ginzburg Lagrangian of a spinor field $\psi$ whose components are $\psi_s = \sqrt{\rho_s} \exp(i\theta_s)$, $s=-S \dots S$. $\mathcal{L}_t$ is the ``minimal'' topological term allowed by symmetries, which generalizes a Chern-Simons coupling in the language of spinors. A factor of $\Phi_0$ is necessary in $\mathcal{L}_t$ if the time-reversal symmetry is to be protected in the absence of a U(1) magnetic field ($\phi_0=0$), and we use anti-commutators (braces) to symmetrize $\mathcal{L}_t$ with respect to the position of $\Phi_0$. We will show that $\eta=\frac{1}{4}$ is real and topologically quantized, making $\mathcal{L}_t$ play a role similar to the Berry's phase.

The equations of motion are expressed in terms of the particle currents
\begin{equation}\label{pCurrents}
j_{\pp\mu} = \epsilon^{\mu\nu\lambda} \partial_{\nu}^{\phantom{a}} \widetilde{j}_{\vv\lambda}^{\phantom{a}}
  \quad , \quad
J_{\pp\mu}^{a} = \epsilon^{\mu\nu\lambda} \partial_{\nu}^{\phantom{a}} \widetilde{J}_{\vv\lambda}^{a}
\end{equation}
defined as curls of the gauge-dependent phase currents:
\begin{eqnarray}\label{Currents}
\widetilde{j}_{\vv\mu} &=& -\frac{i}{2}\Bigl\lbrack\psi^{\dagger}\Phi_0(\partial_{\mu}\psi)
    -(\partial_{\mu}\psi^{\dagger})\Phi_0\psi\Bigr\rbrack \\
\widetilde{J}_{\vv\mu}^{a} &=& -\frac{i}{2}\Bigl\lbrack\psi^{\dagger}\gamma^{a}\Phi_0(\partial_{\mu}\psi)
    -(\partial_{\mu}\psi^{\dagger})\Phi_0\gamma^{a}\psi\Bigr\rbrack \nonumber \ .
\end{eqnarray}
This definition guaranties that $j_{\pp\mu}$ and $J_{\pp\mu}^a$ are gauge-invariant and transform as physical charge ($j_0\to j_0$, $j_i\to -j_i$) and spin ($J_0^a\to -J_0^a$, $J_i^a\to J_i^a$) currents under time-reversal, $\psi^{\phantom{*}}_s\to\psi^*_{-s}$, $\phi_0\to-\phi_0$. In contrast, the phase currents of the $\psi$ field (\ref{Currents}) have inverted parities under time-reversal, implying that $\psi$ is not an ordinary bosonic matter but rather a vortex field. The full relationship between particle and vortex dynamics is established by duality mappings \cite{Dasgupta1981, Fisher1989, Sachdev1990}, which goes beyond the scope of this paper. We will merely emphasize that the present vortex field theory formulation is necessary in order to directly expose the \emph{particle} drift current kinematics in a topological stationary action. We will later explore some physical consequences using this complete description of dynamics. The equations of motion for drift particle currents follow from $\epsilon^{\mu\nu\lambda}(\partial_{\nu}-i\mathcal{A}_{\nu})(\partial_{\lambda}-i\mathcal{A}_{\lambda}) = \epsilon^{\mu\nu\lambda}\partial_{\nu}\partial_{\lambda}-i\Phi^{\mu}$ and the stationary action principle:
\begin{eqnarray}\label{EqMot2}
&& \psi^{\dagger}\frac{\partial\mathcal{L}_{t}}{\partial\psi^{\dagger}} = -\frac{i\eta}{2}\,\psi^{\dagger} \bigl\lbrace
  \partial_{\mu}-i\mathcal{A}_{\mu} , \Phi_{0} \bigr\rbrace
  \nonumber \\ && ~~~~~~~~ \times \Bigl(\epsilon^{\mu\nu\lambda}\partial_{\nu}\partial_{\lambda}-i\Phi^{\mu}\Bigr)\psi+h.c.=0 \ .
\end{eqnarray}
Clearly, the field configurations that satisfy
\begin{equation}\label{EqMot2b}
\psi^{\dagger}\Phi_{0}\Bigl(\epsilon^{\mu\nu\lambda}\partial_{\nu}\partial_{\lambda}
  -i\Phi^{\mu}\Bigr)\psi=0
\end{equation}
also satisfy (\ref{EqMot2}). We again assume the global spin U(1) symmetry and choose the representation $\Phi_\mu^a = \Phi_\mu^z \delta_{az}^{\phantom{a}}$. Defining $\Gamma_{n} = \langle \psi^{\dagger} (\gamma^z)^n \psi \rangle$ at the action saddle-point in incompressible states, we find
\begin{eqnarray}\label{EqMot3}
\!\!\!\!\!\! j_{\pp\mu}^{\phantom{z}} &=& \phi_{0}^{\phantom{z}}\phi_{\mu}^{\phantom{z}}\Gamma_{0}^{\phantom{z}}
  + (\Phi_{0}^z\phi_{\mu}^{\phantom{z}}+\phi_{0}^{\phantom{z}}\Phi_{\mu}^z)\Gamma_{1}^{\phantom{z}}
  + \Phi_{0}^z\Phi_{\mu}^z\Gamma_{2}^{\phantom{z}} \\[0.03in]
\!\!\!\!\!\! J_{\pp\mu}^{z} &=& \phi_{0}^{\phantom{z}}\phi_{\mu}^{\phantom{z}}\Gamma_{1}^{\phantom{z}}
  + (\Phi_{0}^z\phi_{\mu}^{\phantom{z}}+\phi_{0}^{\phantom{z}}\Phi_{\mu}^z)\Gamma_{2}^{\phantom{z}}
  + \Phi_{0}^z\Phi_{\mu}^z\Gamma_{3}^{\phantom{z}} \ . \nonumber
\end{eqnarray}
Let us now recall that the quantum mechanical equation of motion (\ref{EqMot}) describes the drift current $\delta j_i$ of a single particle whose ``density'' is $\delta j_0 = 1$, while (\ref{EqMot3}) describes the current $j_{\pp i}$ and density $j_{\pp 0}$ of many particles (averaged in the saddle-point state of the field theory). Therefore, we wish to compare $\delta j_i$ in (\ref{EqMot}) with $j_{\pp i} / j_{\pp 0}$ in (\ref{EqMot3}). We can immediately see that making replacements $\langle (\gamma^z)^n \rangle \to \Gamma_n$ in the quantum average of (\ref{EqMot}) gives rise to the expression for $\langle \delta j_\mu \rangle$ that agrees in detail with $j_{\pp 0}^{-1} j_{\pp\mu}^{\phantom{1}}$ obtained from (\ref{EqMot3}). Analogous correspondence between equations of motion is found for spin currents by inserting a factor of $\gamma^z$ in (\ref{EqMot2b}). Therefore, $\mathcal{L}_t$ captures the kinematics of drift currents in the combined U(1) and SU(2) ``electromagnetic'' fields. Note that additional solutions allowed by (\ref{EqMot2}) correspond to fluctuations that are suppressed in quantum Hall liquids.

\section{Basic topological orders}

The Landau-Ginzburg part of this theory applied to bosons (microscopically formulated on a lattice) describes superfluid and Mott insulator phases, whose transition is driven by phase $\theta_s$ fluctuations of the spinor components. If the fluctuations respected the U(1)$^{2S+1}$ symmetry, the dynamics near the transition would be captured by $2S+1$ copies of the quantum XY model. The superfluid to Mott-insulator transition can be viewed as the condensation of quantized vortices $\psi$ according to the duality transformation of this model \cite{Dasgupta1981, Fisher1989, Sachdev1990}. Particles are mobile and coherent in the superfluid state, while vortices are gapped and localized into a vortex lattice if their density is finite. A Mott insulator is a dual reflection of the superfluid where particles and vortices exchange their behavior.

We can qualitatively view quantum Hall states as ``arrested'' Mott transitions in which both particles and vortices are abundant and mobile, yet uncondensed and controlled by the cyclotron scales. Duality allows simultaneous mobility of both particles $j_{\pp\mu}$ and vortices $j_{\vv\mu}$ only if vortices are ``attached'' to particles to prevent relative motion. In a quantum Hall state we must imagine that superfluid correlations are not locally lost, but particles have begun localizing so their wavefunctions must acquire vorticity due to the external flux. When every particle becomes a microscopic ``cyclotron'' vortex, it experiences a Magnus force (dual to the Lorentz force) from the residual local phase coherence, which is captured by the topological term $\mathcal{L}_t$.

Armed with this duality picture, we can envision characteristic field configurations $\psi$ in quantum Hall states, which must contain a finite density of singularities in order to take advantage of $\mathcal{L}_t$ (note that any smooth configuration of $\psi$ makes $\mathcal{L}_t$ vanish). Amplitude fluctuations $\rho_s = \langle \psi^\dagger_s \psi^{\phantom{\dagger}}_s \rangle$ are frozen in an incompressible state, while phases may fluctuate freely as long at their winding along any infinitesimal space-time loop $dC$ is quantized as an integer:
\begin{equation}
n_{s}=\frac{1}{2\pi}\oint\limits _{dC}\dd l_{\mu}\,\partial_{\mu}\theta_{s} \in \mathbb{Z} \ .
\end{equation}
Let $dA_{\mu}$ be the vector of the oriented surface element bounded by $dC$; it is perpendicular to the element, with magnitude equal to the element's infinitesimal area. Then, $Q = j_{\pp\mu}dA_{\mu}$ is either the physical charge contained within area $dA_{0}$ for $dA_{\mu} = dA_{0} \, \delta_{\mu,0}$, or the charge pushed through a line segment $dl_i$ in a time interval $dt$ for $dA_{\mu} = dl_i dt \, \delta_{\mu,i}$. We can similarly extract the amount of spin $S^z = J_{\pp\mu}^zdA_{\mu}^{\phantom{z}}$, and use (\ref{Currents}) to obtain:
\begin{eqnarray}\label{FracExc}
Q &=& \oint\limits _{dC}\dd l_{\mu}\,\widetilde{j}_{\vv\mu} =
      \sum_{s=-S}^S 2\pi n_{s} (\phi_0^{\phantom{z}}+s\Phi_0^z) \rho_{s} \\
S^{z} &=& \oint\limits _{dC}\dd l_{\mu}\,\widetilde{J}_{\vv\mu}^{z} =
      \sum_{s=-S}^S 2\pi n_{s} (\phi_0^{\phantom{z}}+s\Phi_0^z) s \rho_s \nonumber \ .
\end{eqnarray}
There are $2S+1$ spinor phases $\theta_s$ for spin-$S$ particles whose vorticities $n_s$ must be quantized. We can characterize a topological ground-state by $n_s = m_s(Q,S^z)$ vorticities attached to a microscopic particle with quantum numbers $(Q,S^z)$. But, only when $m_s(Q,S^z) = m_s \delta_{s,S^z}$ we can solve (\ref{FracExc}) to find $\rho_s$ that are independent of $(Q,S^z)$:
\begin{equation}\label{rhoS}
\rho_s = \frac{1}{2\pi m_s (\phi_0^{\phantom{z}} + s\Phi_0^z)} \ .
\end{equation}
Therefore, a ground-state is defined by $2S+1$ integers $m_s$ (restricted by $\rho_s>0$) whose reciprocals generalize the concept of filling factors. Fractional excitations carry quantum numbers generated by all combinations of $n_s\in\mathbb{Z}$ in (\ref{FracExc}):
\begin{equation}
\delta Q = \sum_s\frac{n_s}{m_s} \quad , \quad \delta S^z = \sum_s\frac{n_s}{m_s}s \ ,
\end{equation}
and are related to the ground-state symmetries by (\ref{EqMot3}). The ground-state charge and spin densities are:
\begin{equation}
j_{\pp 0} = \sum_{s=-S}^{S}\frac{\phi_0^{\phantom{z}}+s\Phi_0^z}{2\pi m_s} \quad , \quad
J_{\pp 0}^z = \sum_{s=-S}^{S}\frac{s\phi_0^{\phantom{z}}+s^2\Phi_0^z}{2\pi m_s} \ .
\end{equation}

For spin $S = \frac{1}{2}$ particles in magnetic field with the filling factor $\nu=2m_{+1/2}^{-1}=2m_{-1/2}^{-1}$, the absence of a spin-orbit coupling $\Phi_0^z=0$ yields the Laughlin sequence $j_{\pp 0} = \nu \phi_0 / 2\pi$, $J_{\pp 0}^z = 0$, with elementary fractional excitations $\delta Q = \nu/2$, $\delta S^z = \pm\nu/4$. We see that spin must be fractionalized just like charge, effectively reducing $\hbar$ by an integer because the $S^a$ operators must obey the Lie algebra. A correlated TR-invariant TI in zero magnetic field $\phi_0=0$ exhibits the same combined spin and charge fractionalization when $m_{+1/2} = -m_{-1/2}$. Charge and spin can be independently fractionalized for $m_{+1/2} \neq \pm m_{-1/2}$, but this generally corresponds to a broken TR-symmetry, even in the zero magnetic field.

Fractionalization is dynamically related to vortex winding numbers, but fractional statistics and topological orders are shaped by the topological term $\mathcal{L}_t$, which is sensitive only to vortex and monopole topological defects of the ``gauge fields'' $b_{s\mu} = \partial_\mu \theta_s$. Let us integrate by parts the left-most derivative of $\mathcal{L}_t$ in (\ref{TopLG}) and write $\mathcal{L}_t = \mathcal{L}_t' + \delta\mathcal{L}_t$, where $\delta\mathcal{L}_t$ is the total derivative of a field bilinear. By Gauss' theorem, $\delta\mathcal{L}_t$ picks monopoles $\partial_\mu (\epsilon^{\mu\nu\lambda} \partial_\nu b_{s\lambda}) \neq 0$ (its space-time integral in the action equals the total monopole charge). However, monopoles can exist (cost a finite energy) only at the system boundaries because $b_{s\mu}$ are phase gradients. We will be interested only in bulk properties below. The bulk $\mathcal{L}_t'$ is sensitive to vortex singularities and yields Chern-Simons (CS) and ``background field'' (BF) effective theories in incompressible states with emergent U(1)$^n$ symmetry, which we will discuss next. The relationship between the spinor field $\psi$ and the CS/BF gauge fields from the examples below is summarized in the Table \ref{CSBF}.

Consider the vicinity of a vortex $\psi_s = \sqrt{\rho_s} \exp(i\theta_s)$, where $\theta_s$ winds $m_s$ times about the flux tube at ${\bf r}_0$. We define phase gauge fields $b_{s\mu} = \partial_\mu \theta_s$ and organize them into a diagonal matrix $B_\mu = \textrm{diag}(b_{s\mu})$ whose flux is $\Phi_B^\mu = \textrm{diag}(\phi_s^\mu)$. As $\rho_s \approx \textrm{const}$, $\phi_s^\mu = \epsilon^{\mu\nu\lambda} \partial_\nu b_{s\lambda}$ vanishes in the plane perpendicular to the tube, except at ${\bf r} = {\bf r}_0$. We assume that important field configurations have a finite density of vortex defects. The phase $\theta_s$ fluctuation kinematics is captured by:
\begin{eqnarray}\label{EffTh}
\mathcal{L}_{t}' &\!\!=\!\!& \frac{i\eta}{2}\Bigl\lbrack(\partial_{\mu}-i\mathcal{A}_{\mu})\psi\Bigr\rbrack^\dagger \Phi_0
  \Bigl\lbrack(\epsilon^{\mu\nu\lambda}\partial_{\nu}\partial_{\lambda}-i\Phi^{\mu})\psi\Bigr\rbrack + \cdots \nonumber \\
&\!\!=\!\!& \frac{i\eta}{2}\textrm{tr}\Bigl\lbrack(B_\mu-\mathcal{A}_\mu)\Phi_0
  (\Phi_B^\mu-\Phi_{\phantom{B}}^\mu)(\psi\psi^\dagger)\Bigr\rbrack + \cdots \ .
\end{eqnarray}
The other symmetrized terms in $\mathcal{L}_{t}'$ denoted by ellipses yield the same kind of the final result. This effective theory still handles many different incompressible quantum liquids on the same footing, but we can derive its simplified special forms in concrete topological ground states characterized by the incompressible densities (\ref{rhoS}). Consider first spin $S=0$ particles that can couple only to U(1) electromagnetic fields. We can immediately recover the standard CS theory of such particles \cite{WenQFT2004} from the $S=0$ representation of (\ref{EffTh}) by substituting (\ref{rhoS}) for $\psi\psi^\dagger$. We will define a CS gauge field $c_\mu = m^{-1} b_\mu$ in order to associate one flux quantum of $c_\mu$ with $m$ windings of $\theta$, that is a single microscopic particle. The particle current (\ref{pCurrents}) becomes
\begin{equation}
j_{\pp\mu} = \frac{1}{2\pi} \epsilon^{\mu\nu\lambda} \partial_\nu c_\lambda \ .
\end{equation}
We must also set the topological coupling to $\eta=\frac{1}{4}$. This yields the standard CS form of $\mathcal{L}_{t}'$ (written in imaginary time):
\begin{equation}
\mathcal{L}_{t}' \to \mathcal{L}_\textrm{CS} = -i \left\lbrack
  -\frac{m}{4\pi}\epsilon^{\mu\nu\lambda}c_\mu\partial_\nu c_\lambda + j_{\pp\mu} a_\mu \right\rbrack \ .
\end{equation}
The value $\eta=\frac{1}{4}$ is physically required by the fermionic (odd $m$) or bosonic (even $m$) statistics of physical microscopic particles that this CS theory implements \cite{WenQFT2004}.

\begin{table}
\centering
\begin{tabular}{|c|c|}
\hline 
$S=0$ quantum Hall & $S=\frac{1}{2}$ quantum spin-Hall \\
\hline
\hline 
$\psi=\frac{1}{\sqrt{2\pi m\phi_{0}}}\, e^{i\theta}$ & $\psi=\frac{1}{\sqrt{\pi m\Phi_{0}^{z}}}\left(\begin{array}{c}
e^{i\theta_{\uparrow}}\\
e^{i\theta_{\downarrow}}\end{array}\right)$ \\
\hline 
$
b_{\mu}=\partial_{\mu}\theta$
 & $B_{\mu}=\left(\begin{array}{cc}
\partial_{\mu}\theta_{\uparrow} & 0\\
0 & \partial_{\mu}\theta_{\downarrow}\end{array}\right)$ \\
\hline 
$c_{\mu}=\frac{1}{m}\partial_{\mu}\theta$ & $\begin{array}{c}
c_{\mu}^{c}=\frac{1}{m}\partial_{\mu}(\theta_{\uparrow}-\theta_{\downarrow})\\[0.05in]
c_{\mu}^{s}=\frac{1}{m}\partial_{\mu}(\theta_{\uparrow}+\theta_{\downarrow})\end{array}$ \\
\hline
\end{tabular}
\caption{\label{CSBF}The relationship between the spinor field $\psi$ and the Chern-Simons $c_\mu$ or BF theory $c_\mu^c, c_\mu^s$ gauge fields. The amplitudes of $\psi$ are frozen in quantum Hall states, while the charge and spin ``phase'' fluctuations are captured by $c_\mu, c_\mu^c$ and $c_\mu^s$ respectively. We assumed the minimal representations of the U(1) and SU(2) symmetry groups, and the presence of TR-symmetry in the latter.}
\end{table}

The effective theory (\ref{EffTh}) written in the $S=\frac{1}{2}$ representation of the U(1)$^2$ symmetry conserves both charge and the $S^z$ spin projection. The gauge fields $\mathcal{A}_\mu$, $B_\mu$ and their fluxes can all be simultaneously diagonalized, while $\psi\psi^\dagger$ becomes a matrix whose off-diagonal elements $\sqrt{\rho_s\rho_{s'}} \exp\lbrack i(\theta_s-\theta_{s'})\rbrack$, $s\neq s'$ do not matter under the trace in (\ref{EffTh}) when all other matrices are diagonal. Assuming a paramagnetic TR-invariant quantum spin-Hall liquid with $m_{\pm 1/2} = \pm m$, we can write $B_\mu = \frac{m}{2} (c_\mu^s + c_\mu^c\sigma^{z})$ to obtain the BF theory \cite{Cho2010, Neupert2011}:
\begin{equation}
\mathcal{L}_{t}' \to  \mathcal{L}_\textrm{BF} = -i \left\lbrack
  -\frac{m}{4\pi}\epsilon_{\phantom{\mu}}^{\mu\nu\lambda}c_{\mu}^{c}\partial_{\nu}^{\phantom{a}}c_{\lambda}^{s}
  +j_{\pp\mu}^{\phantom{z}}a_{\mu}^{\phantom{a}}+J_{\pp\mu}^{z}A_{\mu}^{z} \right\rbrack \ .
\end{equation}
We used the gauge $\mathcal{A}_{0}=-\epsilon_{ij}x_{i}\Phi_{j}$, $\mathcal{A}_{i}=-\frac{1}{2}\epsilon_{ij}x_{j}\Phi_{0}$ for convenience, and expressed the particle charge and spin currents (\ref{pCurrents}) in terms of the gauge fields:
\begin{equation}
j_{\pp\mu} = \frac{1}{2\pi} \epsilon_{\phantom{\mu}}^{\mu\nu\lambda} \partial_{\nu}^{\phantom{a}} c_{\lambda}^{c}
  \quad , \quad
J_{\pp\mu}^{z} = \frac{1}{4\pi} \epsilon_{\phantom{\mu}}^{\mu\nu\lambda} \partial_{\nu}^{\phantom{a}} c_{\lambda}^{s} \ .
\end{equation}
The transverse spin conductivity is $\sigma_{xy}^s = J_{\pp i}^z/\phi_i = (2\pi m)^{-1}$ in the units $\hbar = e = 1$.

\section{Generalizations and applications}

We have demonstrated so far that the effective theory (\ref{TopLG}) can describe Laughlin quantum Hall states shaped by external magnetic fields or $S^z$-conserving spin-orbit couplings. Since the character of the external fields and the nature of their coupling to matter are formally determined by the symmetry group and its representation, (\ref{TopLG}) can capture any combination of SU($N$) quantum Hall effects. This merely requires reinterpreting $\gamma^a$, $a \in (1,2,\dots,N^2-1)$ as SU($N$) generators in some representation. Furthermore, (\ref{TopLG}) can be easily extended to describe any hierarchical quantum Hall state. The main physical feature of hierarchical states is the presence of multiple emergent flavors of low-energy quasiparticles that participate in the quantum Hall liquid. The quasiparticle flavor acts as a conserved quantum number associated with an emergent low-energy symmetry. A CS theory associates a separate gauge field to each flavor state, and couples them via the ``K-matrix'' to define their mutual exchange statistics \cite{WenQFT2004}. We can obtain an equivalent structure at the level of (\ref{EffTh}) by enlarging the spinor symmetry group to $\lbrack$U(1)$\times$SU(2)$\rbrack^n$ and using a general linear relationship between the particle currents and the curls of vortex currents that extends (\ref{pCurrents}) to multiple flavor states. A detailed discussion of this procedure will be presented elsewhere. Here, we wish to emphasize that all CS and BF theories studied so far in the topological insulator literature \cite{Santos2011} are restricted to the implementations of the U(1)$^{2n}$ symmetry group in the present theory.

The main benefit of the effective topological Lagrangian $\mathcal{L}_t$ in (\ref{TopLG}) is that it is not limited to quantum spin-Hall states like the CS theory, but can also describe incompressible quantum liquids shaped by the SU(2) gauge fields like (\ref{Bernevig}) that do not conserve spin. Such fractional topological liquids without an analogue in quantum Hall states may be possible to obtain in the Rashba spin-orbit-coupled TIs, and will be studied elsewhere. Their topological orders are guaranteed to be captured by $\mathcal{L}_t$ due to its large symmetry that covers all possible configurations of the external flux gauge fields. In other words, the validity of (\ref{TopLG}) in these general cases is justified by its gauge symmetry. Note that our justification of $\mathcal{L}_t$ via the equations of motion was possible only by specializing to the quantum (spin) Hall states, which have topologically protected drift (edge) currents and whose symmetry is restricted to powers of $U(1)$. Since the SU(2) gauge fields of the Rashba spin-orbit coupling are fundamentally non-Abelian, the topologically ordered states that they create likely exhibit novel kinds of non-Abelian fractional statistics. Most generally, $\mathcal{L}_t$ can describe topological orders in any representation of any continuous gauge symmetry group. Note, however, that the written form of $\mathcal{L}_\textrm{LG}$ in (\ref{TopLG}) can be justified only in quantum Hall states, but can be easily amended to properly describe conventional states.

Realistic systems have ``perturbations'' that violate the SU(2) gauge symmetry of any finite-$S$ representation. This is not a priori detrimental to topological order, but may gap out edge states \cite{Levin2009} and spoil the spin-Hall effect even in ideally $S^z$-conserving models. Also, the quantum numbers of fractional quasiparticles depend on the conserved quantities. Charge is always conserved and can be fractionalized, while ``fractional spin'' has to be understood in general as a quasiparticle's degree of freedom derived from the electron's spin that transforms non-trivially under TR. It becomes a quantum number only if $S^z$ is conserved, or if no perturbations spoil the symmetries of the ideal Rashba spin-orbit-coupled model (\ref{Bernevig}).

Being very general, the constructed topological field theory can describe any two-dimensional system whose topological state is created by continuum magnetic fields and/or spin-orbit couplings with non-zero average fluxes. This includes all available two-dimensional TIs, which are uncorrelated but may or may not exhibit spin conservation in the absence of disorder and other perturbations. There are presently unavailable but feasible systems that could exhibit fractionalized versions of spin-orbit topological states without spin conservation, and the presented theory is the first proposal of a theory that could capture their properties. One promising system are ultracold atoms in artificial gauge fields that we mentioned in the introduction. Another is the heterostructure featuring a Bi$_2$Se$_3$ quantum well embedded between a superconductor and a gated conventional insulator, which can in principle host superconducting and insulating topological liquids of $p$-wave Cooper pairs created by the superconducting proximity effect and dynamically empowered by the spin-orbit coupling \cite{Nikolic2011a, Nikolic2012a}. This is an example of the system whose description requires a non-commutative SU(2) gauge field in the spin $S=1$ representation that the proposed theory provides.

Since the proposed effective theory (\ref{TopLG}) naturally admits non-commuting symmetry groups, one may ask if it is capable of describing the well-known non-Abelian quantum Hall states, such as the Moore-Read pfaffian state \cite{Moore1991}. While the full answer to this question is not presently known, it is likely positive because suitable extensions of (\ref{TopLG}) can reproduce at least some of the effective CS gauge theories that have been proposed for the pfaffian states \cite{Fradkin1998, Fradkin2001}. For example, the fermionic pfaffian state with the filling factor $\nu=1/q$ might be described by a U(1)$_q \times$O(2)$_{2q}$ CS Lagrangian, where the charge and neutral sectors of fractional quasiparticle excitations are governed by the level $q$ U(1) and the level $2q$ O(2) CS theories respectively \cite{Fradkin2001}. Such a ``double'' CS theory can be obtained as the low-energy limit of an extended (\ref{TopLG}), which contains a complex scalar field $\psi$ for the charge sector, and a two-component real spinor field $\phi = (\phi_1, \phi_2)$ for the neutral sector. The desired incompressible quantum liquid can exist if the fluctuations of $|\psi|^2$ and $|\phi|^2 = \phi_1^2 + \phi_2^2$ are frozen, so that the remaining fluctuations can be expressed in terms of the appropriate U(1) and O(2) CS gauge fields. The O(2) group contains the SO(2)$\cong$U(1) subgroup, which admits infinitesimal gauge transformations and thus a continuum-limit gauge-theory formulation. However, O(2) also contains a Z$_2$ transformation (reflection) that cannot be carried out infinitesimally. Consequently, the theory (\ref{TopLG}) should be regularized and formulated on a lattice in order to enable the desired U(1)$_q \times$O(2)$_{2q}$ CS low-energy description (just like a Z$_2$ gauge theory \cite{senthil00}). It is this extra Z$_2$ transformation that makes the O(2) gauge fields non-commutative, and gives rise to the non-Abelian statistics through the neutral sector. A more detailed exploration of these issues is left for future work.

\section{Conclusions}

In conclusion, the topological term of (\ref{TopLG}) describes topological orders of general incompressible quantum liquids shaped by externally applied magnetic and/or spin-orbit fields in the continuum limit. Not all such liquids are quantum Hall states in non-Abelian ``magnetic'' fields, but they generally feature quasiparticles whose fractional statistics is captured by the topological term.

\section{Acknowledgments}

I am very grateful to Michael Levin for insightful discussions, and to the Aspen Center for Physics for its hospitality. This research was supported by the Office of Naval Research (grant N00014-09-1-1025A), the National Institute of Standards and Technology (grant 70NANB7H6138, Am 001), and the U.S. Department of Energy, Office of Basic Energy Sciences, Division of Materials Sciences and Engineering under Award DE-FG02-08ER46544 (summer 2011).

\section{References}




\end{document}